\documentclass[aps,preprint,nofootinbib,superscriptaddress]{revtex4}

\usepackage{amssymb}

\usepackage{epsfig}
\usepackage[bookmarksnumbered,bookmarksopen,colorlinks,citecolor=blue,linkcolor=blue]{hyperref}

\begin{document}

\title{ Distinguishing fission-like events from deep-inelastic collisions }

\author{Hong Yao}
\affiliation{ School of Physics, Beihang University, Beijing 102206, People's Republic of China}
\affiliation{ Department of Physics, Guangxi Normal University, Guilin 541004, People's Republic of
China }

\author{Cheng Li}
\affiliation{ Department of Physics, Guangxi Normal University, Guilin 541004, People's Republic of
China }
\affiliation{ Guangxi Key Laboratory of Nuclear Physics and Technology, Guilin 541004, People's Republic of
China }

\author{Houbing Zhou}
\affiliation{ Department of Physics, Guangxi Normal University, Guilin 541004, People's Republic of
China }
\affiliation{ Guangxi Key Laboratory of Nuclear Physics and Technology, Guilin 541004, People's Republic of
China }

\author{Ning Wang}
\email{wangning@gxnu.edu.cn}\affiliation{ Department of Physics,
Guangxi Normal University, Guilin 541004, People's Republic of
China }
\affiliation{ Guangxi Key Laboratory of Nuclear Physics and Technology, Guilin 541004, People's Republic of
China }
\begin{abstract}

We propose two functions to distinguish fission-like events from quasi-elastic (QE) scattering and deep inelastic collisions (DIC), for a better analysis of the measured mass-total kinetic energy  distributions of binary fragments formed in fusion-fission reactions. We note that the ratio of capture to DIC events evidently decreases with the decreasing of the depth of the capture pocket predicted from the Skyrme energy density functional, with which the capture pocket could be extracted from the measured mass-energy distributions. Together with the improved quantum molecular dynamics simulations, in which the typical contact time of the reaction partners is smaller than 200 fm/c for QE and is larger than 600 fm/c for fission-like events, we find that the ratio of capture to touching cross section systematically increases with the pocket depth.

\end{abstract}
\maketitle

\begin{center}
\textbf{I. INTRODUCTION}
\end{center}

Synthesis of super-heavy nuclei (SHN) through heavy-ion fusion reactions is
a field of very intense studies in the recent decades \cite{Hof00,Hof04,Mori04a,Ogan06,Ogan10,Ogan15,Ogan22,Sob,Mori20,Pomo18,Adam04,Wang15}.
For the fusion systems leading to the synthesis of SHN, the estimation of the optimal excitation energy and the evaporation residue (EvR) cross sections is vital, since the measurements are very time-consuming and the EvR cross sections could be as small as a few femtobarn \cite{Wang11,Adam20,Zhang23,Nov20} for the reactions producing 119 and 120 elements. In the practical calculations, the reaction process leading to the synthesis of SHN can be divided into two or three steps, in which the fusion probability $P_{\rm CN}$  after capture is the most unclear part. The quasi-fission (QF) wherein the composite system fails to evolve into a compound nucleus (CN) after capture and breaks apart before reaching compact equilibrium shapes, significantly complicates the description of fusion process. The mass-energy distributions (MEDs) and mass-angle distributions (MADs) of fragments for a number of reactions have been measured to explore the competition between fusion and QF in the formation of heavy and SHN \cite{Toke85,Shen87,Hind92,Riet11,Riet13,Koz14,Koz16,Koz19,Albe20,Hind21,Sen22}. To evaluate the value of $P_{\rm CN}$, the contribution of fragments with masses $A_{\rm CN}/2 \pm  20$ u is usually measured \cite{Koz19} considering that the mass distributions of the CN-fission fragments for the systems with $Z\sim108 - 114$ have the symmetric Gaussian shape with a standard deviation of about 20 u according to the liquid drop model, with which the relative contributions of symmetric fragments in the capture cross sections (as a sum of QF, CN-fission and EvR cross section, i.e., $\sigma_{\rm cap}=\sigma_{\rm QF}+\sigma_{\rm FF}+\sigma_{\rm ER}$) can be obtained. It is there necessary to firstly distinguish the capture events from the quasi-elastic (QE) and deep-inelastic collisions (DICs) fragments in the mass-energy distributions as clearly as possible.

For medium-mass fusion-fission reactions, the fission-like fragments can be unambiguously separated from the mass-total kinetic energy (M-TKE) distributions, since the contribution of fragments originate from QF processes is negligible, and the reaction products located between the quasi-elastic peaks are assumed as totally relaxed events, i.e., as fission-like fragments. However, for the reaction systems leading to the synthesis of SHN, especially those with very shallow capture pockets such as $^{86}$Kr+$^{198}$Pt, it is difficult to clearly distinguish the QF and DIC since both the processes are binary with full momentum transfer, in which the composite system separates in two main fragments without forming a CN and are characterized by sufficient energy dissipation and mass transfer. For example, the M-TKE distributions for four composite systems with  $Z = 114$ are analyzed in Ref. \cite{Koz19}. We note that the red contour lines  (within the events selected as capture ones) in the M-TKE distributions for $^{86}$Kr+$^{198}$Pt are quite different from those for the three others. Even for the same reaction $^{48}$Ca+$^{244}$Pu, the contour lines in \cite{Koz14} are different from those in \cite{Koz19} at the overlapping region for QF and DIC events. It is therefore necessary to propose a uniform method to distinguish the fission-like events from the QE and DICs, with which the capture cross sections could be unambiguously measured.

On the other hand, according to dinuclear system (DNS) model, the fusion process of super-heavy system is described as the evolution of the DNS in which nucleons are transferred from the light nucleus to the heavy one \cite{Adam97}. The initial DNS is localized in the minimum of the capture pocket of the nucleus-nucleus potential $V(R)$, where the sufficient energy dissipation and mass transfer between fragments take place. The depth of the capture pocket (which is also defined as the quasi-fission barrier height $B_{\rm qf}$ in \cite{Wang11}) significantly influences the fusion probability. For light and medium-mass fusion reactions, the value of $B_{\rm qf}$ is high enough to make QF an improbable decay mode at incident energies close to the Coulomb barrier. For the reactions between two massive nuclei, the capture process could become difficult when the capture pocket disappears. It is known that the distribution for the fusion (capture) barrier heights can be experimentally obtained from the measured fusion (capture) excitation functions or back-angle quasi-elastic scattering excitation functions \cite{Timm97,Zhang10}. How to extract the depth of the capture pocket from the measured MEDs for heavy system is an interesting question. In this work, we attempt to distinguish the fission-like fragments from the QE scattering and DICs in MEDs. Simultaneously, we would like to investigate the relationship between the relative yields of DIC events and the capture pocket depth.

\begin{center}
\textbf{ II. SYSTEMATIC ANALYSIS OF THE MEDS }\\
\end{center}

It is known that the fission-like fragments are located between the quasi-elastic peaks which can be unambiguously determined as mentioned previously. Together with the energy conservation condition, one can obtain the TKE maximum of fission-like fragments. The total energy of the compound nuclei should be equal to the total energies of the binary fragments if neglecting the energies of the evaporated particles and the residue excitation energies of the fragments, i.e.,
\begin{eqnarray}
E_{\rm CN}+E_{\rm CN}^*=E_1+E_2+{\rm TKE}.
\end{eqnarray}
Here, $E_{\rm CN}$, $E_1$ and $E_2$ denotes the energies of the CN and binary fragments at their ground states, respectively. $E_{\rm CN}^*$ denotes the excitation energy of the CN. According to Eq.(1), one obtains the upper limit of the TKE for symmetric fission, $Q_s+E_{\rm CN}^*$. $Q_s=E (A_1+A_2 ,Z_1+Z_2 )-2E (A_m,Z_m)$ denotes the reaction $Q$-value for symmetric fission, with $A_1$, $A_2$, $Z_1$ and $Z_2$ denoting the mass and charge numbers of the projectile and target nuclei. $A_m=(A_1+A_2)/2$ and $Z_m=(Z_1+Z_2)/2$ denote the mass and charge number of symmetric fission fragments, respectively. The reaction $Q$-value for unmeasured super-heavy system is from the prediction of the Weizs\"acker-Skyrme (WS4) mass
model \cite{WS4} with which the known masses can be reproduced with an rms error of $\sim0.3$ MeV \cite{Zhao22} and the known $\alpha$-decay energies of SHN can be reproduced with deviations smaller than 0.5 MeV \cite{Ogan15,Wang15}.

\begin{figure}
\includegraphics[angle=0,width=1\textwidth]{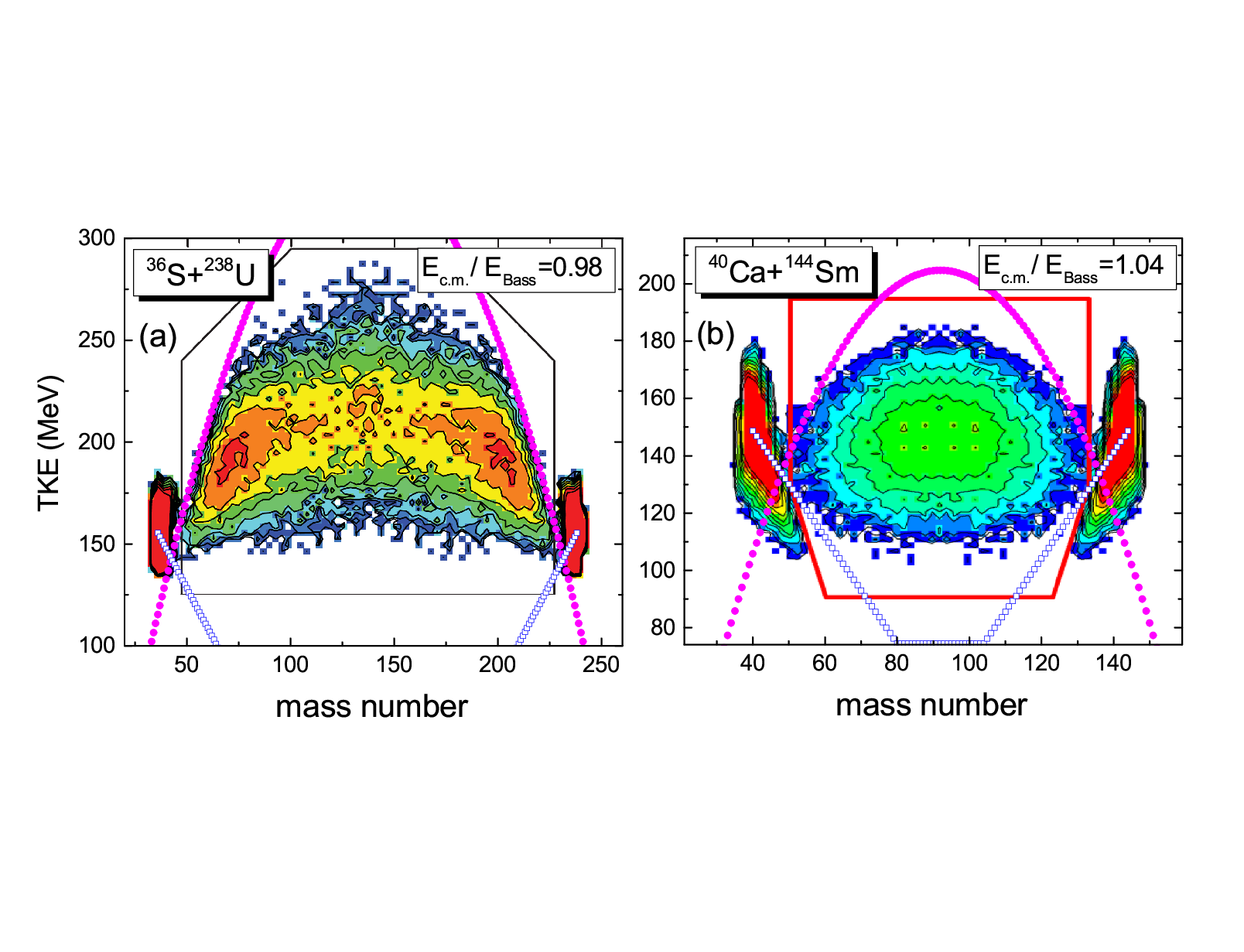}
\caption{(Color online)  Mass - total kinetic energy distributions for the reactions $^{36}$S+$^{238}$U and $^{40}$Ca+$^{144}$Sm at energies around the Bass barrier. The pink curves and the blue squares denote the results from Eq.(2) and Eq.(3), respectively.
\\
Source: Adapted from Ref. \cite{Itkis11,Bog21}.  }
\end{figure}

With the upper limit $Q_s+E_{\rm CN}^*$ for symmetric fission, we propose an inverted parabolic curve to describe the upper limit of the TKE of fission-like fragments,
\begin{eqnarray}
F_1=(Q_s+E_{\rm CN}^*)-(Q_s+E_{\rm CN}^*-Q_0)\frac{(A-A_m)^2}{(A_1-A_m)^2}.
\end{eqnarray}
Here, $Q_0$ denotes the reversed value of $Q$ at the entrance channel. In addition, we also propose a function to distinguish the DIC and fission-like fragments,
\begin{eqnarray}
F_2= \left\{
\begin{array} {r@{\quad:\quad}l}
  E_{\rm c.m.}(1-|A-A_1|/80)  &   A_1\leqslant A < A_m    \\
  E_{\rm c.m.}(1-|A-A_2|/80)  &   A_m\leqslant A \leqslant A_2    \\
\end{array} \right.
\end{eqnarray}
together with a truncation for the value of $F_2$, i.e., $E_{\rm c.m.}/2\leqslant F_2\leqslant E_{\rm c.m.}$.  From the DICs of $^{208}$Pb+$^{110}$Pd at an incident energy of $18\%$ above the Bass barrier, the energy loss in nucleon transfer was studied in Ref. \cite{Rehm81}. It is observed that the energy loss per exchanged proton approaches $\sim 3$ MeV at TKE loss larger than $\sim 80$ MeV, which is generally consistent with the coefficient ($E_{\rm c.m.}/80\approx 3\pm 1$ MeV) adopted in Eq.(3) for the fusion-fission reactions considered in this work.

With $F_1$ and $F_2$, the measured M-TKE distributions are separated into three regions. In Fig. 1, we show the M-TKE distributions for the reactions $^{36}$S+$^{238}$U and $^{40}$Ca+$^{144}$Sm at energies around the Bass barrier. The black lines in (a) and the red ones in (b) denote the contour lines in the Ref. \cite{Itkis11} and \cite{Bog21}, respectively, and the events within the contour lines are selected as the fission-like events. The pink circles and the blue squares denote the results from Eq.(2) and Eq.(3), respectively. One can see that almost all fission-like fragments are located in the region $F_2 < {\rm TKE} \leqslant F_1$, and the QE events are located outside of $F_1$. In Fig. 2, we show the M-TKE distributions for $^{48}$Ca+$^{238}$U, $^{48}$Ca+$^{248}$Cm, $^{52}$Cr+$^{248}$Cm and $^{64}$Ni+$^{238}$U. The fission-like fragments produced in these different reaction systems can all be clearly distinguished with the uniform region $F_2 < {\rm TKE} \leqslant F_1$. For $^{52}$Cr+$^{248}$Cm and $^{64}$Ni+$^{238}$U, we note that the lower part of the contour lines given in Ref. \cite{Nov20} are almost the same as $F_2$. For $^{48}$Ca+$^{238}$U, the fragments with $100<{\rm TKE}< 150$ MeV and mass number about $A_m\pm 50$ (outside the red contour lines) are also selected as the fission-like fragments according to Eq.(3).

In addition, we also note from Fig. 2 that for these heavy reaction systems, there exists a number of events with large energy dissipation but relative small mass transfer, i.e., located in the region ${\rm TKE}\leqslant F_1$ and ${\rm TKE}\leqslant F_2$. In the DICs of $^{208}$Pb+$^{94}$Zr, $^{208}$Pb+$^{110}$Pd, $^{208}$Pb+$^{148}$Sm and $^{208}$Pb+$^{170}$Er  at incident energies close to the Coulomb barrier, the authors also observed that a few tens of nucleons are transferred between the reaction partners together with quite large TKE loss (up to $100\sim 200$ MeV) \cite{Rehm81}. Considering the DIC characteristics, the events located in the region ${\rm TKE}\leqslant F_1$ and ${\rm TKE} \leqslant F_2$ are defined as the DIC events in this work.

\begin{figure}
\includegraphics[angle=0,width=0.9\textwidth]{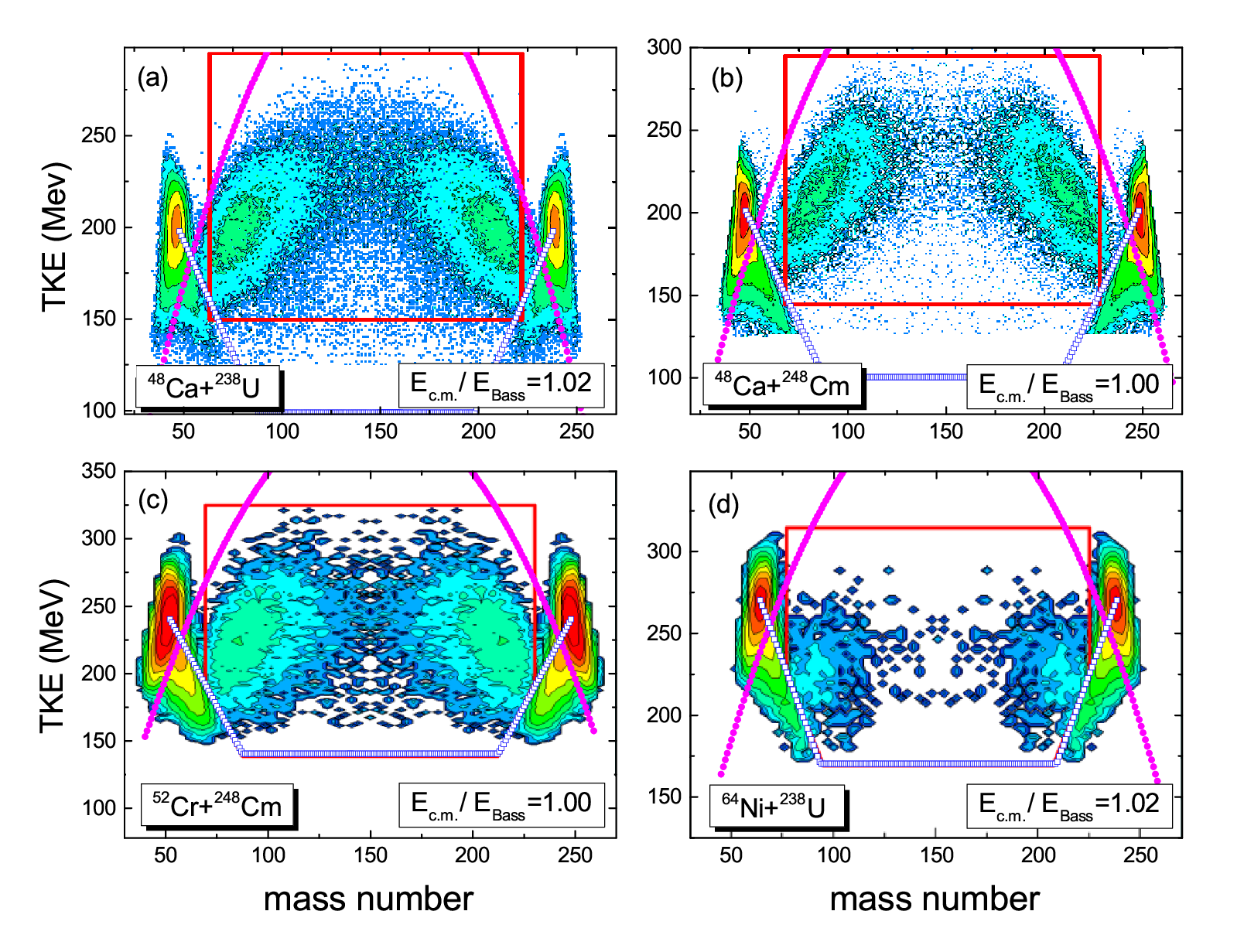}
\caption{(Color online) The same as Fig. 1, but for $^{48}$Ca+$^{238}$U, $^{48}$Ca+$^{248}$Cm, $^{52}$Cr+$^{248}$Cm and $^{64}$Ni+$^{238}$U.
\\
Source: Adapted from Ref. \cite{Koz14,Nov20}.  }
\end{figure}

To test the validity of $F_2$ for distinguishing QF and DIC, we show the M-TKE distributions for $^{58}$Fe+$^{208}$Pb and $^{86}$Kr+$^{208}$Pb in Fig. 3 and $^{48}$Ca+$^{244}$Pu, $^{48}$Ti+$^{238}$U, $^{52}$Cr+$^{232}$Th and $^{86}$Kr+$^{198}$Pt in Fig. 4. For $^{58}$Fe+$^{208}$Pb, the boundaries (with abrupt change of the yields) for the reaction products located between the quasi-elastic peaks are relatively evident. One sees that the boundaries can be well reproduced by the blue squares according to Eq.(3). Except for the two reactions induced by $^{86}$Kr in which the capture pockets generally disappear (will be discussed later), the boundaries along the abrupt change of the yields can all be clearly described by $F_2$. With uniform contour lines to distinguish QF and DIC, the systematic behavior of the capture cross sections could be more accurately analyzed.

\begin{figure}
\includegraphics[angle=0,width=0.9\textwidth]{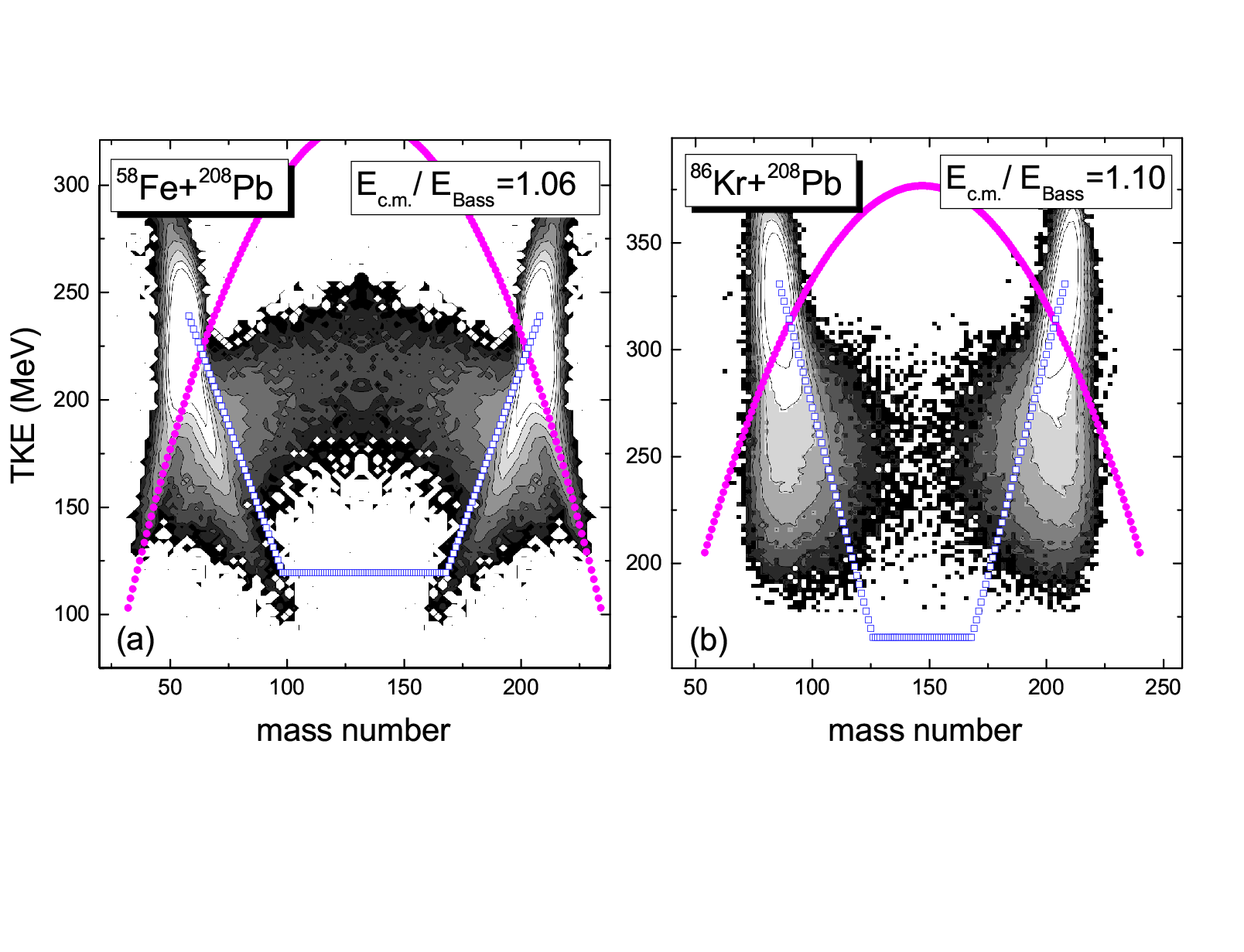}
\caption{(Color online) The same as Fig. 1, but for $^{58}$Fe+$^{208}$Pb and $^{86}$Kr+$^{208}$Pb.
\\
Source: Adapted from Ref. \cite{Itkis02}. }
\end{figure}

\begin{figure}
\includegraphics[angle=0,width=0.9\textwidth]{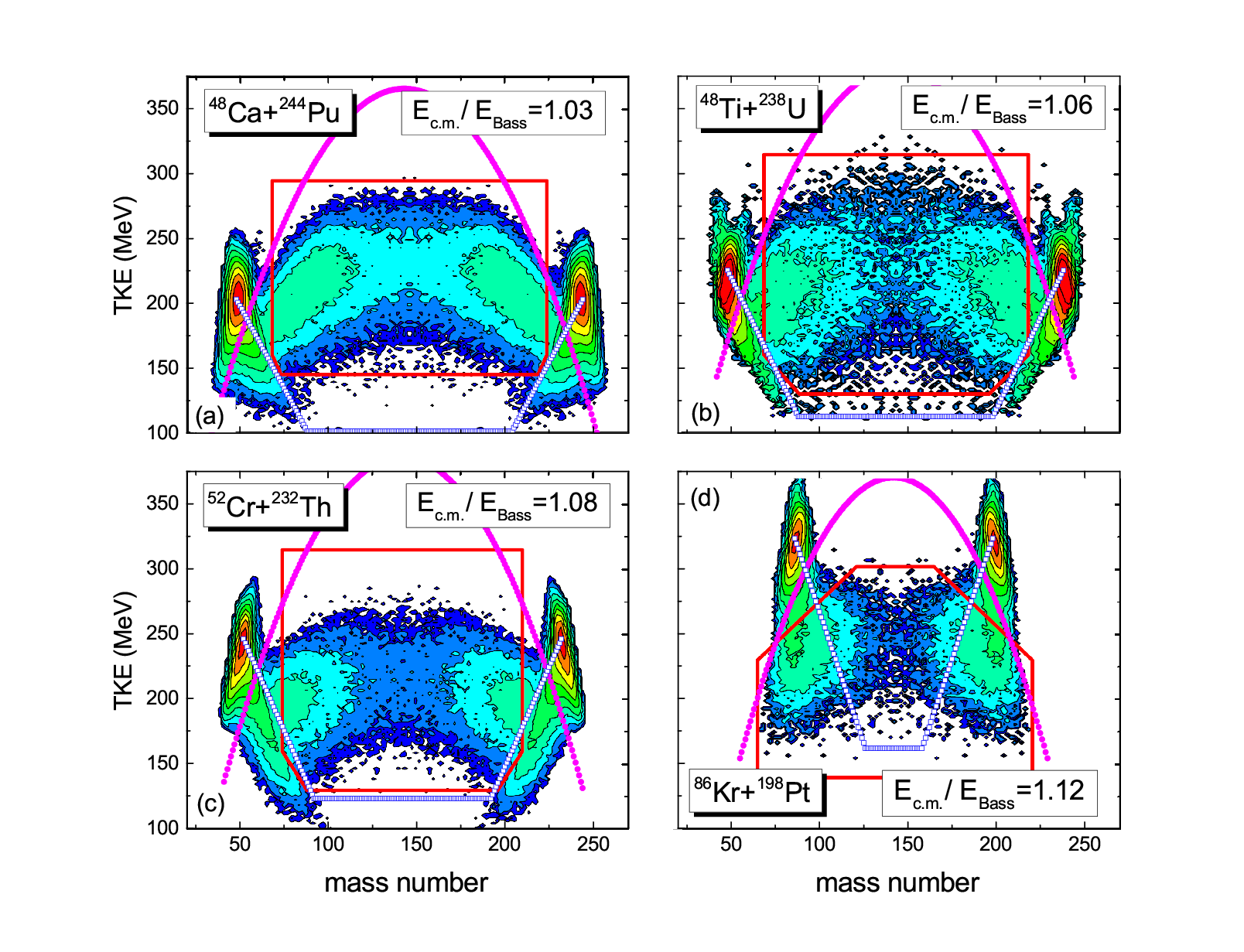}
\caption{(Color online) The same as Fig. 1, but for $^{48}$Ca+$^{244}$Pu, $^{48}$Ti+$^{238}$U, $^{52}$Cr+$^{232}$Th and $^{86}$Kr+$^{198}$Pt.
\\
Source: Adapted from Ref. \cite{Koz19}.  }
\end{figure}

\begin{center}
\textbf{ III. DEPTH OF THE CAPTURE POCKET }\\
\end{center}

From the comparison of Fig. 1 and Fig. 3, one sees that the yields of DIC fragments significantly increase in $^{58}$Fe+$^{208}$Pb and $^{86}$Kr+$^{208}$Pb. To understand the physics behind, we investigate the capture pocket in the nucleus-nucleus potential based on the Skyrme energy density functional (EDF) \cite{Vau72, Brack,Bart02} with the parameter set SkM*\cite{Bart82}.

The entrance-channel nucleus-nucleus potential $V(R)$ between the reaction partners can be expressed as \cite{Deni02,liumin}
\begin{eqnarray}
V(R) = E_{\rm tot}(R) - E_{\rm proj} - E_{\rm targ},
\end{eqnarray}
where $R$ is the center-to-center distance between two fragments. $E_{\rm tot}(R)$ denotes the total energy of the nuclear system, $E_{\rm proj}$ and $E_{\rm targ}$ denote the ground state energies of the projectile and target nuclei, respectively. The total energy of a nuclear system can be expressed as the integral of the Skyrme EDF ${\mathcal H}(\bf{r})$,
\begin{eqnarray}
E_{\rm tot}(R) =  \int \; {\mathcal H}[ \rho_{1p}({\bf
r})+\rho_{2p}({\bf r}-{\bf R}), \rho_{1n}({\bf r})+\rho_{2n}({\bf
r}-{\bf R})] \; d{\bf r}, \nonumber
\end{eqnarray}
\begin{eqnarray}
E_{\rm proj} = \int \; {\mathcal H}[ \rho_{1p}({\bf r}), \rho_{1n}({\bf
r})] \;
d{\bf r}, \\
E_{\rm targ} = \int \; {\mathcal H}[ \rho_{2p}({\bf r}), \rho_{2n}({\bf
r})] \; d{\bf r}.
\end{eqnarray}
Here, $\rho_{1p}$, $\rho_{2p}$, $\rho_{1n}$ and $\rho_{2n}$ are the frozen proton and neutron densities of the projectile and
target described by spherical symmetric Fermi functions. In the calculations of $E_{\rm tot}$, $E_{\rm proj}$, $E_{\rm targ}$, and the densities of the reaction partners, the same Skyrme EDF  is adopted combining the extended Thomas-Fermi (ETF2) approach for describing the kinetic energy density and the spin-orbit density in the EDF (see Ref. \cite{liumin} for details). Under frozen density approximation, there exists a capture pocket in $V(R)$ around projectile-target touching configuration. In addition to the frozen barrier height $B_0$, the depth of the pocket which is defined as the quasi-fission barrier height $B_{\rm qf}$ can also be obtained \cite{Wang11}.

\begin{figure}
\includegraphics[angle=0,width=0.8\textwidth]{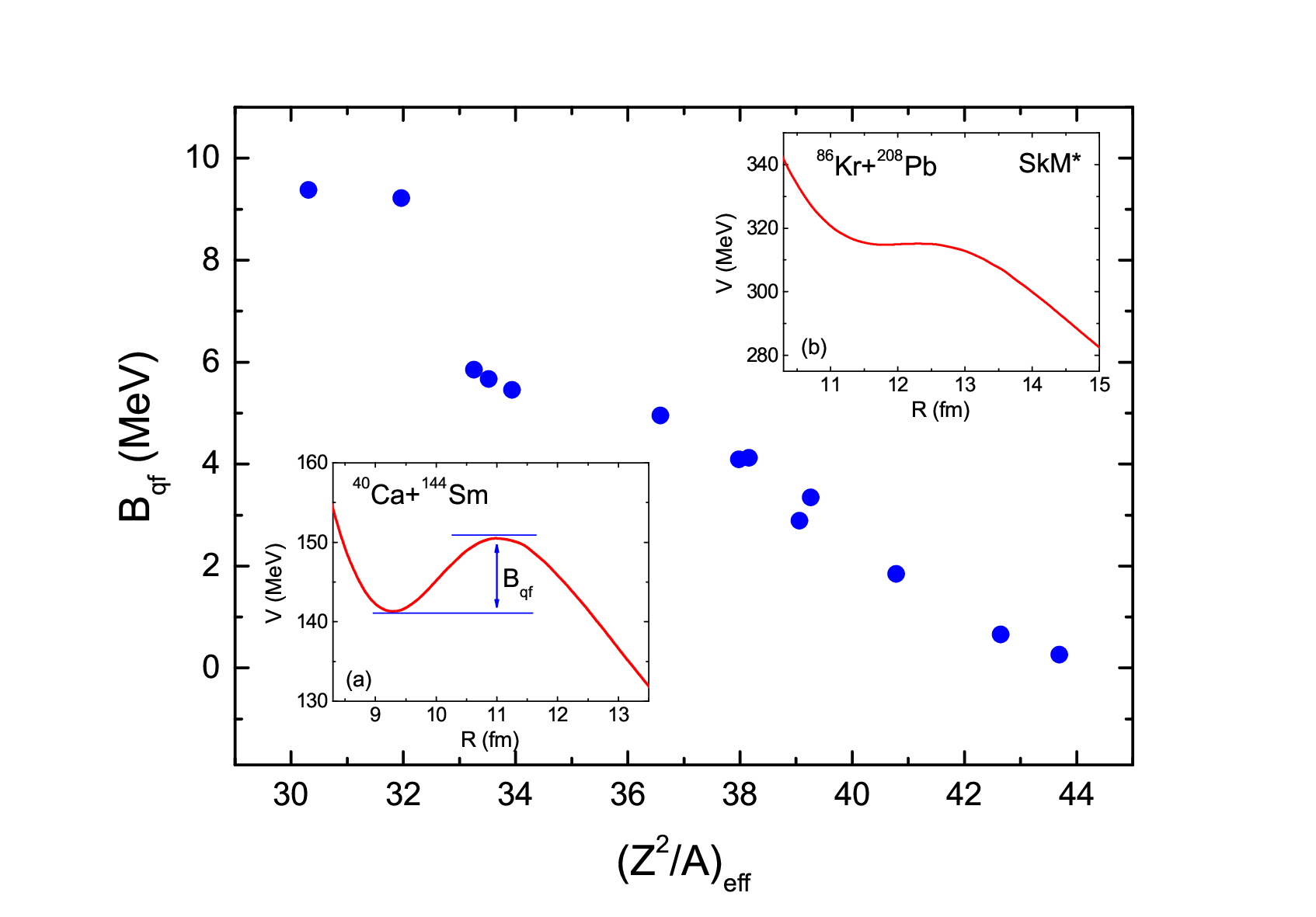}
\caption{(Color online) Capture pocket depth as a function of effective fissility. }
\end{figure}

The obtained values of $B_{\rm qf}$ for 13 reaction systems are listed in Table I. Simultaneously, the corresponding effective fissility \cite{Swiat82}
\begin{eqnarray}
(Z^2/A)_{\rm eff}=\frac{4 Z_1 Z_2}{A_1^{1/3}A_2^{1/3}(A_1^{1/3}+A_2^{1/3})}
\end{eqnarray}
are also listed in the table. The capture pocket depths for these systems are shown in Fig. 5 as a function of effective fissility. One sees that the values of $B_{\rm qf}$ systematically decrease with the increasing of effective fissility. It is thought that for the systems with $(Z^2/A)_{\rm eff}>33 $ the influence of quasi-fission becomes evident and an extra push is needed to achieve fusion \cite{Swiat82,BOCK82}. For $^{36}$S+$^{238}$U and $^{40}$Ca+$^{144}$Sm, the depth of the capture pocket are larger than 9 MeV and the corresponding values of effective fissility are smaller than 33, which implies that the QF events in these two reactions could be few. From Fig. 1, we note that the yields at the overlapping region for the QF and DIC fragments (along $F_2$ and far from QE) are really few as expected, which is consistent with the estimation from the calculated capture pocket depth. From Fig. 2 to Fig. 4, one sees that with the decreasing of $B_{\rm qf}$, the yields of fission-like fragments systematically decrease. Especially for $^{64}$Ni+$^{238}$U in Fig. 2(d), the yields of fission-like fragments are evidently smaller than those for $^{52}$Cr+$^{248}$Cm, with a decrease of $B_{\rm qf}$ by 1.5 MeV. For $^{86}$Kr+$^{198}$Pt and $^{86}$Kr+$^{208}$Pb, the yields of DIC fragments are much larger than those of fission-like fragments, since the depths of the capture pockets approach zero. It implies that the relative yields of DIC fragments could be used to probe the depth of the capture pocket for fusion-fission reactions.

\begin{table}
\caption{ The entrance channel properties for the reactions under study: $E_{\rm Bass}$ and $B_{\rm qf}$  denote the Bass barrier \cite{Bass74} and the quasi-fission barrier height, respectively. $(Z^2/A)_{\rm eff}$ denotes the effective fissility \cite{Swiat82}.}
\begin{tabular}{ccccc}
 \hline\hline

 ~~~reaction~~~  & ~~~$E_{\rm Bass}$ (MeV)~~~ & ~~~$Q$ (MeV)~~~ & ~~~$B_{\rm qf}$ (MeV)~~~ & ~~~$(Z^2/A)_{\rm eff}$  ~~~   \\
\hline
 $^{36}$S+$^{238}$U   &  158.5   &  -115.4      &  9.38   &	 30.3    \\
 $^{40}$Ca+$^{144}$Sm &  143.2   &  -105.8      &  9.22   &  32.0    \\
 $^{48}$Ca+$^{238}$U  &  193.8   &  -160.8      &  5.85   &  33.3    \\
 $^{48}$Ca+$^{244}$Pu &  197.3   &  -163.9      &  5.67   &  33.5    \\
 $^{48}$Ca+$^{248}$Cm &  201.0   &  -169.6      &  5.46   &  33.9    \\
 $^{48}$Ti+$^{238}$U  &  213.8   &  -170.9      &  4.95   &  36.6    \\
 $^{58}$Fe+$^{208}$Pb &  226.1   &  -205.0      &  4.09   &  38.0    \\
 $^{52}$Cr+$^{232}$Th &  227.7   &  -187.4      &  4.12   &  38.2    \\
 $^{52}$Cr+$^{248}$Cm &  240.5   &  -204.0      &  3.35   &  39.3    \\
 $^{64}$Ni+$^{208}$Pb &  241.3   &  -224.3      &  2.89   &  39.1    \\
 $^{64}$Ni+$^{238}$U  &  265.5   &  -238.6      &  1.85   &  40.8    \\
 $^{86}$Kr+$^{198}$Pt &  288.7   &  -280.6      &  0.66   &  42.6    \\
 $^{86}$Kr+$^{208}$Pb &  301.5   &  -302.9      &  0.26   &  43.7    \\

 \hline\hline
\end{tabular}
\end{table}

To study the relative yields of DIC fragments, we define the sum of capture cross section and DIC cross section as the touching cross section, i.e.,
\begin{eqnarray}
\sigma_T=\sigma_{\rm cap}+\sigma_{\rm DIC}.
\end{eqnarray}
To investigate the ratio $\sigma_{\rm cap}/\sigma_T$, we use the improved quantum molecular dynamics (ImQMD) model \cite{ImQMD14,ImQMD17,Li20} to simulate the capture and DIC processes. In Ref. \cite{Wu19}, the multinucleon transfer reaction $^{58}$Ni+$^{124}$Sn at energies around the Coulomb barrier is investigated by using the time dependent Hartree-Fock calculations. It is find that the contact time for QE scattering at an incident energy of $\sim 0.94 E_{\rm Bass}$ is about 150 fm/c for head-on collisions. In the ImQMD simulations for $^{136}$Xe+$^{198}$Pt at an incident energy of $\sim 1.10 E_{\rm Bass}$, Li et al. note that the contact time of the QE events is mainly in the ranges of $t_{\rm con} \leqslant 200$ fm/c \cite{Li20} which is consistent with the TDHF calculations.

\begin{figure}
\includegraphics[angle=0,width=0.8\textwidth]{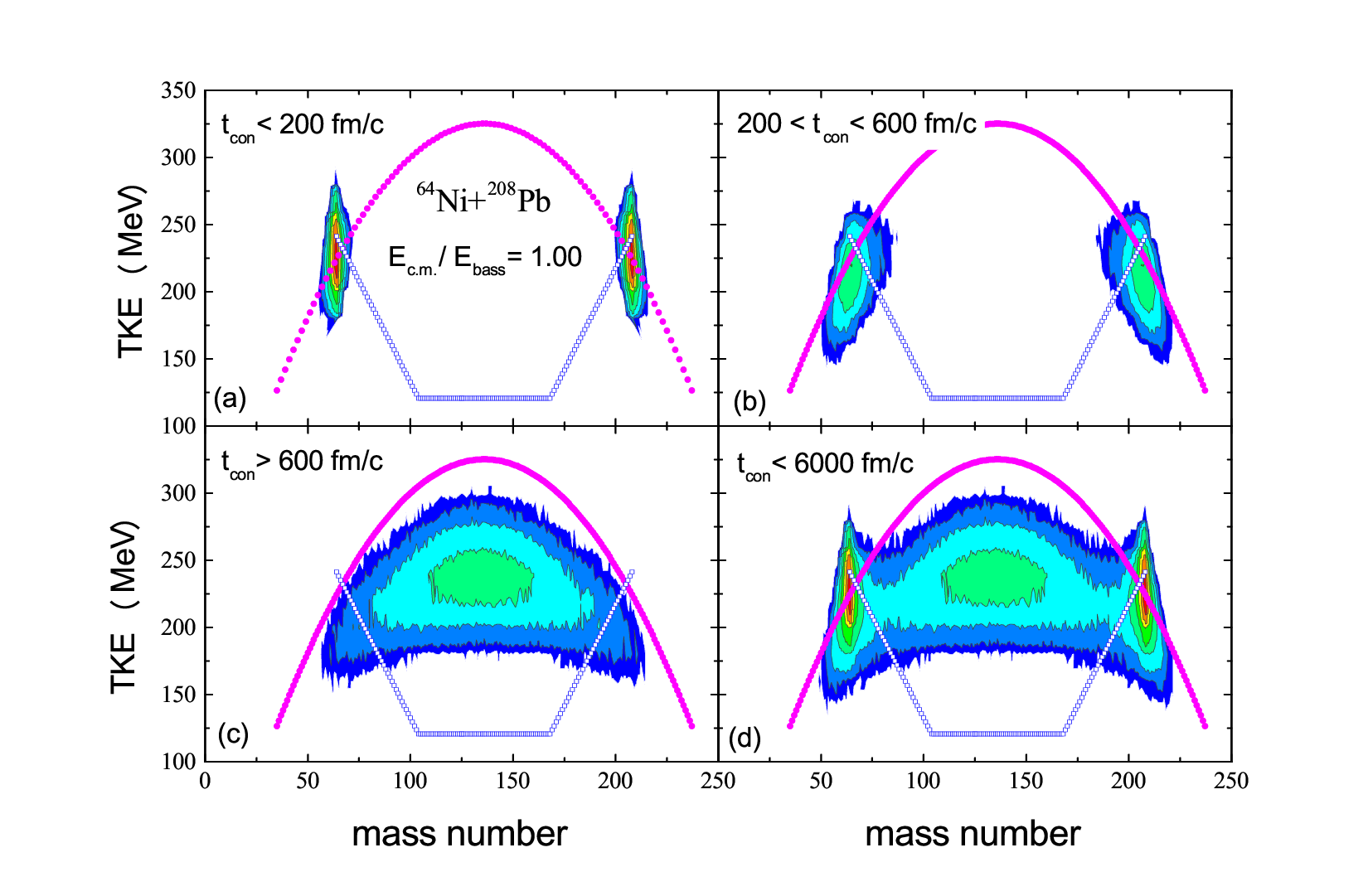}
\caption{(Color online) M-TKE distributions for the events with different contact time from the ImQMD simulations for head-on collisions of $^{64}$Ni+$^{208}$Pb. The pink curves and the blue squares denote the results from Eq.(2) and Eq.(3), respectively.}
\end{figure}

In Fig. 6, we show the M-TKE distributions for head-on collisions of $^{64}$Ni+$^{208}$Pb at an incident energy of Bass barrier by using the ImQMD model. One sees that the fragments produced in the collisions with contact time of $200 < t_{\rm con} < 600$ fm/c are mainly located at the DIC region and the fission-like fragments are mainly from the events with contact time larger than 600 fm/c. It is thought that QF takes place within a few zeptoseconds ( $1 zs = 10^{-21}s$) \cite{Toke85,Shen87,Hind92,Riet11,Riet13}. For example, the average QF contact time extracted from the measured QF mass-angle distributions  for $^{64}$Ni+ $^{238}$U \cite{Albe20} and $^{86}$Kr+$^{197}$Au \cite{Sen22} at energies around the capture barriers is about 3 zs. It is therefore reasonable to select the events with contact time $ t_{\rm con} \geqslant 600$ fm/c as the capture and $ t_{\rm con} < 200$ fm/c as the QE events in the calculations.


\begin{figure}
\includegraphics[angle=0,width=0.7\textwidth]{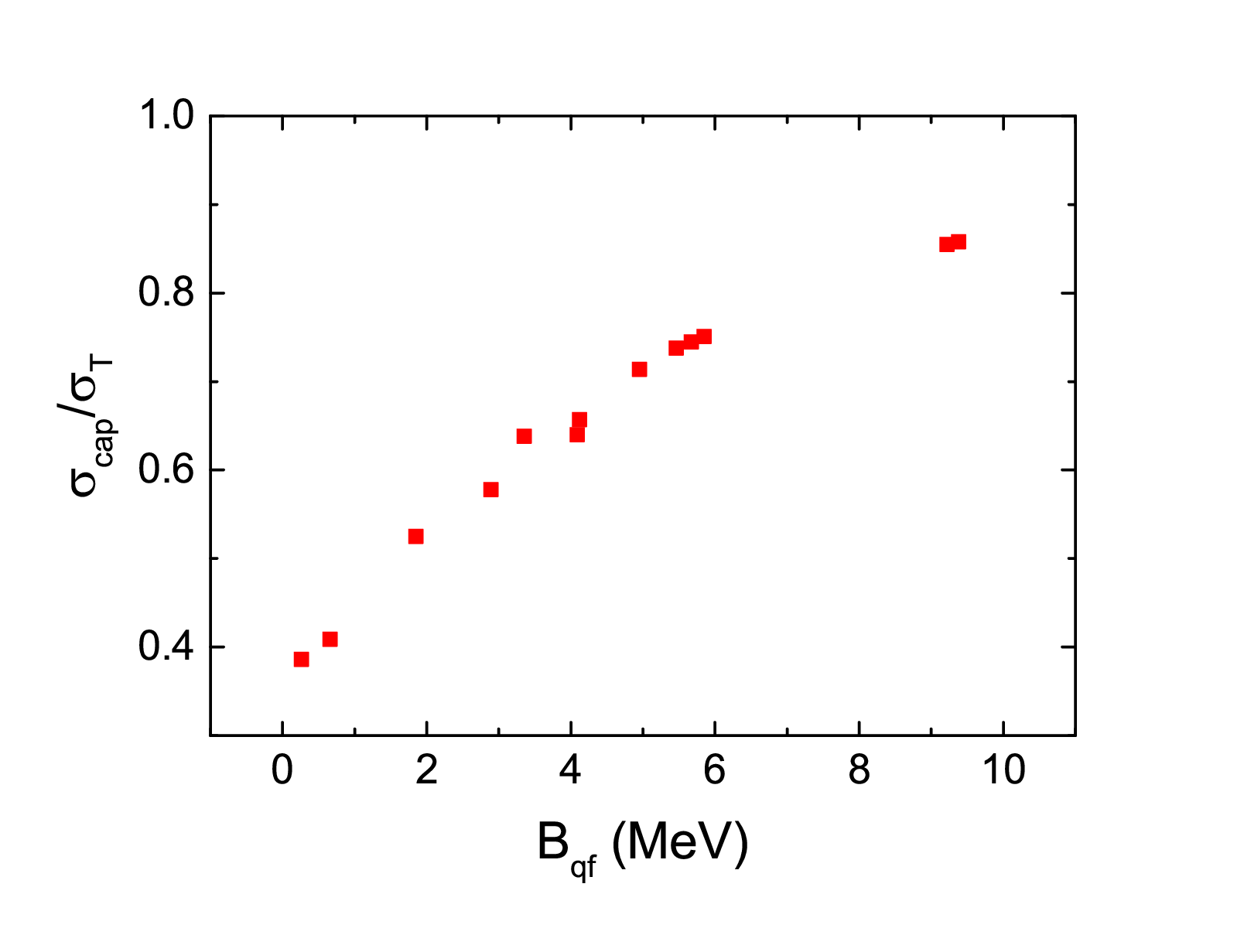}
\caption{(Color online) Calculated ratio $\sigma_{\rm cap}/\sigma_T$ as a function of $B_{\rm qf}$.}
\end{figure}

Through creating 100,000 bombarding events at each incident energy $E_{\rm c.m.}$ and each impact parameter $b$, and counting the number of capture events, one obtains the capture probability $g_{\rm cap}(E_{\rm c.m.},b)$. The corresponding capture cross section can be calculated with
\begin{equation}
\sigma _{\rm cap}(E_{\rm c.m.})=2\pi \int b \, g_{\rm cap} \, db
\simeq 2\pi \sum b \, g_{\rm cap} \, \Delta b.
\end{equation}
Here, we set $\Delta b=1$ fm and the maximal impact parameter $b_{\rm max}=10$ fm. The events with the contact time $ t_{\rm con} \geqslant 600$ fm/c are selected in the calculation of the capture probability, as mentioned above. For calculations of the touching cross section, we use the similar procedure but selecting the events with $ t_{\rm con} \geqslant 200$ fm/c.

In Fig. 7, we show the calculated ratio $\sigma_{\rm cap}/\sigma_T$ at an incident energy of Bass barrier. One sees that the ratio $\sigma_{\rm cap}/\sigma_T$ systematically increases with the depth of capture pocket as expected. For $^{86}$Kr+$^{208}$Pb the calculated ratio is 0.39, and the probability of DIC is larger than that of fission-like events which is generally consistent with the observation from Fig. 3(b). For $^{40}$Ca+$^{144}$Sm with a value of $B_{\rm qf}=9.22$ MeV, the ratio rises up to 0.86 and the DIC cross sections are much smaller than the capture cross sections, which implies that the touching cross section roughly equals to the capture (fusion) cross section for light fusion systems since the contributions of DIC and QF are negligible at energies around the Coulomb barrier. Very recently, the fusion-evaporation reaction $^{40}$Ca+$^{175}$Lu with a value of $B_{\rm qf}=9.21$ MeV has been studied at the gas-filled recoil separators SHANS and SHANS2 \cite{Zhang24}. It is found that the influence of QF in the analysis of evaporation residue cross sections of this reaction can be neglected. The calculated ratios $\sigma_{\rm cap}/\sigma_T$ from ImQMD indicate that the relative yield of DIC fragments is a sensitive observable to probe the depth of the capture pocket for fusion-fission reactions.

\begin{center}
\textbf{IV. SUMMARY}
\end{center}

We propose two formulaes to distinguish the fission-like events from quasi-elastic (QE) scattering and deep inelastic collisions (DIC) for the measured mass-total kinetic energy (TKE) distributions of fragments in fusion-fission reactions. Through considering the upper limit of TKE from the reaction $Q$-value and the energy loss per particle of about 3 MeV, the fission-like events can be clearly distinguished from QE and DIC for the 12 fusion-fission reactions under study. In addition, we note that the ratio of capture to DIC events evidently decreases with the decreasing of the depth of the capture pocket obtained from the Skyrme energy density functional together with the extended Thomas-Fermi (ETF2) approach and frozen density approximation. For $^{36}$S+$^{238}$U and $^{40}$Ca+$^{144}$Sm, the depth of the capture pocket are larger than 9 MeV and the yields of DIC fragments are relatively few, while for $^{86}$Kr+$^{198}$Pt and $^{86}$Kr+$^{208}$Pb, the yields of DIC fragments are much larger than those of fission-like fragments, since the capture pockets almost disappear. Together with the improved quantum molecular dynamics simulations, in which the typical contact time of the reaction partners is smaller than 200 fm/c for QE scattering and is larger than 600 fm/c for the fission-like events, we find that the ratio of capture to touching cross section systematically increases with the depth of the capture pocket. It indicates that the relative yields of DIC fragments could be used to probe the depth of the capture pocket for fusion-fission reactions.

\begin{center}
\textbf{ACKNOWLEDGEMENTS}
\end{center}
This work was supported by National Natural Science Foundation of
China (Nos. 12265006, 12365016, U1867212), Guangxi Natural Science Foundation (Nos. 2023GXNSFBA026008, 2023GXNSFAA026016, 2017GXNSFGA198001). N. W. is grateful to Huiming Jia for valuable discussions.

\end{document}